\documentclass[oldversion]{aa}
\usepackage{graphicx}
\usepackage{natbib}
\usepackage{hyperref}
\usepackage{longtable}
\usepackage{lscape}
\usepackage{amsmath}
\usepackage{amssymb}
\begin{document}

   \title{Radial velocity survey of spatially resolved young, low-mass binaries
}


   \author{Stephen Durkan\inst{1,2} \and
          Markus Janson\inst{2} \and
          Simona Ciceri\inst{2} \and
          Wolfgang Brandner\inst{3} \and
          Joshua Schlieder\inst{4} \and
          Thomas Henning\inst{3} \and
          Micka{\"e}l Bonnefoy\inst{5} \and
          Juliet Kankare\inst{1} \and
          Christopher A. Watson\inst{1} 
	  }

   \offprints{Stephen Durkan}

   \institute{Astrophysics Research Center, Queens University Belfast, Belfast, Northern Ireland, UK\\
	\email{sdurkan01@qub.ac.uk, j.kankare@qub.ac.uk, c.a.watson@qub.ac.uk}
	\and
             Department of Astronomy, Stockholm University, Stockholm, Sweden\\
              \email{markus.janson@astro.su.se, simona.ciceri@astro.su.se}
         \and
	    Max Planck Institute for Astronomy, Heidelberg, Germany\\
	\email{brandner@mpia.de, henning@mpia.de}
	\and
	    NASA Exoplanet Science Institute, Caltech, Pasadena, California, USA\\
	\email{jschlied@ipac.caltech.edu}
	\and
	    Univ. Grenoble Alpes, IPAG, Grenoble, France\\
	\email{mickael.bonnefoy@univ-grenoble-alpes.fr}
             }

   \date{Received ---; accepted 08 June 2018}

   \abstract{The identification and characterisation of low-mass binaries is of importance for a range of astrophysical investigations. Low-mass binaries in young ($\sim10 - 100 $ Myr) moving groups (YMGs) in the solar neighborhood are of particular significance as they provide unique opportunities to calibrate stellar models and evaluate the ages and coevality of the groups themselves. Low-mass M-dwarfs have pre-main sequence life times on the order of $\sim100$ Myr and therefore are continually evolving along a mass-luminosity track throughout the YMG phase, providing ideal laboratories for precise isochronal dating, if a model-independent dynamical mass can be measured. AstraLux lucky imaging multiplicity surveys have recently identified hundreds of new YMG low-mass binaries, where a subsample of M-dwarf multiples have estimated orbital periods less than 50 years. We have conducted a radial velocity survey of a sample of 29 such targets to complement the astrometric data. This will allow enhanced orbital determinations and precise dynamical masses to be derived in a shorter timeframe than possible with astrometric monitoring alone, and allow for a more reliable isochronal analysis. Here we present radial velocity measurements derived for our sample over several epochs. We report the detection of the three-component spectroscopic multiple 2MASS J05301858-5358483, for which the C component is a new discovery, and forms a tight pair with the B component. Originally identified as a YMG member, we find that this system is a likely old field interloper, whose high chromospheric activity level is caused by tidal spin-up of the tight BC pair. Two other triple systems with a tight pair exist in the sample, 2MASS J04244260-0647313 (previously known) and 2MASS J20163382-0711456, but for the rest of the targets we find that additional tidally synchronized companions are highly unlikely, providing further evidence that their high chromospheric activity levels are generally signatures of youth. }


\titlerunning{Radial Velocity Survey of Low-Mass Binaries}
\authorrunning{S. Durkan et al.}

   \maketitle
%

\section{Introduction}
\label{sec:intro}

Low-mass stars in multiple systems are increasingly playing an important role in stellar astrophysics. Statistically constraining their multiplicity characteristics and population properties provides clues on their formation and evolutionary pathways \citep[e.g.,][]{Burgasser2007,Bate2012,Duch2013}, potentially connecting higher-mass stars to brown dwarfs \citep[e.g.,][]{Luhman2007, Chabrier2014}. The identification and characterisation of low-mass multiples is also highly relevant to several fields of exoplanet study. For example, direct imaging surveys typically exclude visual binaries \citep[e.g.,][]{Laf2007, Janson2011, Vigan2012, Rameau2013,Vigan2017} due to the intricacies of planet detection within the combined point-spread function (PSF) patterns of multiple stars, alongside the formation and long-term stability barriers faced by any possible planet orbiting such systems. Recently however, efforts have also been made to probe the population of wide separation circumbinary planets through dedicated imaging studies of stellar multiples \citep[e.g.,][]{Thalmann2014, Bonavita2016}. Binary identification aids target selection for such studies. 

Also of fundamental importance to any imaging study is a good estimate of target age. This is critical for estimating the mass or initial entropy of planets and brown dwarfs using mass-luminosity evolutionary models. Such objects are maximally hot, and therefore luminous, directly after formation and gradually cool --- becoming less luminous as the planet ages \citep{Baraffe2003, Burrows2003, Fortney2008}.  Imaging surveys therefore typically target stars in young moving groups (YMGs) \citep[e.g.,][]{Biller2013,Brandt2014,Durkan2016}, young ($\sim10 - 100 $ Myr) co-moving associations of stars in the solar  neighborhood originating from a common birth cluster  \citep[e.g.,][]{Zuckerman2004,Torres2008}. YMGs also provide age estimates of stars that are potentially much more reliable than any accessible technique for individual stars, translating to a more reliable and precise mass estimate of any imaged companion. 

However, uncertainties remain in the ages of YMGs. For instance, the age estimates in the AB Dor \citep[e.g.,][]{Torres2008, Barenfeld2013, Bell2015} and USco \citep{deZeeuw1999,Pecaut2012, Song2012} associations vary by more than a factor of two between the lowest and highest reasonable estimates in the literature. These age uncertainties, although relatively small compared to field star ages, dominate the mass uncertainty of an imaged sub-stellar companion and can lead to disparity in the mass estimates of analogous studies \citep[e.g., J010335;][]{Delorme2013, Janson2017}, potentially furthering ambiguity to its interpretation as a planet or brown dwarf. A better understanding of YMG ages is essential for more robust constraints to be placed on companion masses.

In this regard, YMG M-dwarf multiples can be very useful. As M-dwarfs have considerably long pre-main sequence lifetimes \citep[$\sim100$ Myr e.g.,][]{ Baraffe1998}, they are continually evolving along a mass-luminosity track throughout the YMG phase, providing ideal laboratories for precise isochronal dating. Such dating analysis can be conducted using spectral properties alone; bolometric luminosity and effective temperature \citep[e.g.,][]{Janson2007}, however, the large degree of uncertainty in the ultra-cool M-dwarf temperature scale introduces a source of systematic error into the analysis. Relating a model independent mass, such as dynamical masses derived from orbitally monitoring M-dwarf multiples, to luminosity allows for a much more robust isochronal analysis. This has been demonstrated for a few YMG binaries \citep[e.g.,][]{ Bonnefoy2009, Kohler2013, Montet2015} however, a broader comprehensive study is of fundamental importance to cover a wider sample of YMGs and to assess dating robustness and coevality within individual YMGs. Model independent M-dwarf masses can also be derived through high precision photometric and spectroscopic observations of double-lined, eclipsing binaries \citep[e.g.,][]{Zhou2014, Zhou2015}. However, such systems are rare and typically much older than any YMG and the $\sim100$ Myr pre-main sequence lifetime.

These arguments for low-mass multiplicity studies and binary characterization have motivated our AstraLux Large M-dwarf Multiplicity surveys, systematic lucky imaging studies of $> 1000$ X-ray active young M-dwarfs, many of which are also high probability YMG members \citep[e.g.,][]{Bergfors2010, Janson2012, Janson2014A, Janson2017}. As one would expect roughly 30\% of these were identified as multiple systems, significantly increasing the number of close YMG M-dwarf binaries. Whilst the ultimate goal of the AstraLux surveys is to derive individual dynamical masses and isochronally date M-dwarf multiples, for improved age constraints on the full YMG population, it is not feasible for the majority of the sample over the current survey lifetime due to the relatively long orbital timescales. However, mass determinations are possible for several systems due to particularly short orbital periods and/or wealth of additional astrometric data in the literature \citep[e.g.,][]{Calissendorff2017}.

However, astrometric monitoring alone is often not enough to derive precise dynamical masses and enable a robust isochronal analysis, a complementary radial velocity (RV) analysis is equally important for a range of purposes. Whilst relative astrometry provides the means to constrain orbital parameters, it is limited to providing the total system mass. An RV analysis provides information about the mass ratio of the system, allowing individual component masses to be derived when coupled with the astrometric information. RV data also provides a third dimension of information, outside the plane of the sky, providing much stronger constraints on mutual orbital parameters (e.g., period, eccentricity, argument of periapsis etc.) in a shorter timeframe than would be possible with either method in isolation \citep{Tuomi2009}.  An additional importance to the RV observations is the ability to efficiently detect further close companions in the system that are unresolved in the images. Identifying such companions is of critical importance for isochronal analysis as any unresolved pair treated as a single star will lead to an incorrect age estimate or model calibration. 

Motivated by this reasoning we have conducted an RV monitoring survey of a sample of 29 high-utility M-dwarf binaries to complement our astrometric observations. As precise mass determinations combining astrometric and RV measurements will only become possible over a several year timescale (in order to sufficiently sample and nearly close orbits) for the majority of the targets, we do not present any individual mass estimates or isochronal analysis here. Instead, we present a spectral analysis of two unique systems that are suspected to host unresolved tidally synchonized companions, and list radial velocities for the sample over several epochs. These measurements will be vital for determining masses over the coming years and achieving the long-term goals of the AstraLux surveys. We also evaluate the likelihood of the presence of additional synchronized companions in the sample.

\section{Target sample}
\label{sec:samp}

In our previous AstraLux M-dwarf multiplicity studies, targets were selected from multiple catalogs of late type stars on the basis of youth; indicated by high-probability YMG membership \citep[e.g.,][]{Malo2013, Malo2014, Kraus2014} and x-ray emission \citep{Lepine2011}.  These studies have detected $> 300$ confirmed M-dwarf binaries, the majority of which were previously undiscovered. \citet{Janson2014B} identified a sub-sample of these binaries that exhibit strong indications of youth and estimated orbital periods less than 50 years. Orbits for these high-utility binaries could be closed on the scale of years to several decades, allowing full orbital parameters and dynamical masses to be constrained within reasonable timeframes. 

The target sample in this study is primarily compiled from the \citet{Janson2014B} sub-sample. We selected 21 targets with orbital periods less than 40 years for complementary RV monitoring. This will allow for full orbital constraints and a robust isochronal analysis to be conducted on a much shorter time-scale than would be possible with astrometric monitoring alone.  We also selected three youthful systems targeted in \citet{Janson2012} that appeared as single stars in the AstraLux images; J042442, J084756 and J232057. However, these targets have either been previously resolved or are spectroscopic binaries, indicating short component separations and therefore rapid orbital periods. As our unresolved images reveal limited astrometric information, radial velocity monitoring of these targets is the most viable method for producing orbital determinations. From the literature we selected an additional five resolved low-mass binary systems that have strong indications of youth and short orbital periods for which additional RV measurements would significantly enhance orbital determinations; J052844 \citep{Janson2007}, J101726 \citep{Bonnefoy2009}, J120727 \citep{Bonavita2016}, J122021\citep{Kohler2001} and J155734 \citep{Laf2014}. Due to the scientific aims of the project, the targets were chosen entirely on individual merit in terms of orbital properties and youth, with no particular considerations regarding sample uniformity in other respects. Our sample is listed in Table \ref{t:1} along with target spectral type, estimated binary orbital period and YMG / association membership. We note that several systems in our sample have previously been identified as members of the Argus association. However, \citet{Bell2015} suggest that Argus is largely contaminated by interlopers and may not represent a single, coeval population. Therefore, it remains unclear if these targets can be associated to any YMG.

\begin{table*}
\centering
\footnotesize
\caption{Target Sample Properties\label{t:1}.}
\begin{tabular}{c c c c c c}
\hline\hline
		2MASS ID &  SpT & Period estimate &  YMG / Association & YMG / Association & YMG / Association\\
		& & (Yr) & & Reference & Age (Myr)\\
		\hline
J01112542+1526214 & M5.0+M6.0 & 12 & $\beta$ Pic & M14 & $24 \pm 3$   \\
J02451431-4344102	& M4.0+M4.5 &	13 & .... & & \\
J02490228-1029220 & M1.5+M3.5+M3.5 & 30 & $\beta$ Pic? & B16 & $24 \pm 3$ \\
J03323578+2843554 & M4.0+M4.5+M5.5 & 8 & $\beta$ Pic & M14, J14 & $24 \pm 3$ \\
J04244260-0647313 & M4.0\footnotemark[1] & $< 0.2$\footnotemark[7] &	Argus & M13 & $40 \pm 10$$^{\,a}$ \\
J04373746-0229282 & M0.0+M3.0 & 29\footnotemark[8]& $\beta$ Pic & M13 & $24 \pm 3$ \\
J04595855-0333123	& M4.0+M5.5 & 9 &	Argus? &	M13, J14 & $40 \pm 10$ \\
J05284446-6526463 & M5.0+M5.5\footnotemark[2] & 1.6\footnotemark[2] & AB Dor & LS06 & $149^{+51}_{-19}\,\rm$ \\
J05301858-5358483 & M3.0+M4.0+M6.0 & 25 & AB Dor** & M14 & $149^{+51}_{-19}\,\rm$ \\
J05320450-0305291 & M2.0+M3.5 & 23 & $\beta$ Pic? & M13, J14 & $24 \pm 3$ \\
J06112997-7213388 & M4.0+M5.0 & 8 &	Carina &	M14 & $45^{+11}_{-7}\,\rm$ \\
J06134539-2352077 & M3.5+M5.0 & 13 & Argus & M14, J14 & $40 \pm 10$  \\
J06161032-1320422 & M3.5+M5.0 & 37 & $\beta$ Pic? & M13, J14 & $24 \pm 3$ \\
J07285137-3014490 & M1.5+M3.5 & 33 & AB Dor & M13 & $149^{+51}_{-19}\,\rm$ \\
J08475676-7854532 & M3.0\footnotemark[3] &	....* & $\eta$ Cha & LM13 & $11\pm 3 $$^{\,b}$ \\
J09075823+2154111	& M2.0+M3.5 &	12 & .... & \\
J09164398-2447428 & M0.5+M2.5 & 10 & .... & \\
J10140807-7636327 & M4.0+M5.5 & 17 & Carina &	M14 & $45^{+11}_{-7}\,\rm$ \\
J10172689-5354265 & M6.0+M6.0\footnotemark[4] & 5.15\footnotemark[4] & $\beta$ Pic &	M13 & $24 \pm 3$ \\
J11315526-3436272 & M2.5+M9.0 &	 5.94\footnotemark[9] &	TW Hya &	M14 &$10 \pm 3$ \\
J12072738-3247002 & M1.0\footnotemark[5] &	4.2\footnotemark[10]\footnotemark[11] & TW Hya &	M14 &$10 \pm 3$ \\
J12202177-7407393 & M1.0\footnotemark[3] & ....*   &	$\epsilon$ Cha & LM13 &$\sim 6$ \\
J13493313-6818291 & M2.0+M4.0+M3.5 & 25 & Argus & M13 & $40 \pm 10$$^{\,a}$ \\
J15573430-2321123 & M1.0\footnotemark[6] &	27\footnotemark[12] &	Upper Scorpius & R15 & $11 \pm 2$$^{\,c}$ \\
J20163382-0711456 & M0.0+M2.0 &	18 &	Argus** &	M14, J14 & $40 \pm 10$$^{\,a}$ \\
J20531465-0221218 & M3.0+M4.0 & 13 &	Argus? &	M13 & $40 \pm 10$$^{\,a}$ \\
J23172807+1936469 & M3.0+M4.5 & 33 &	$\beta$ Pic? & M13, J14 & $24 \pm 3$ \\
J23205766-0147373 & M4.0\footnotemark[1] & ....*  &	Argus &	M14 & $40 \pm 10$$^{\,a}$ \\
J23495365+2427493 & M3.5+M4.5 & 38 & $\beta$ Pic / Columba? &	M13, J14  & $24 \pm 3$ / $42^{+6}_{-4}\,\rm$ \\
\hline
\label{observations}
\end{tabular}
\newline
\footnotetext{}{

\textbf{NOTE} -- Individual spectral types derived by \citet{Janson2012} following the methods of \citet{Daemgen2007}, unless otherwise noted. Estimated orbital period based on system mass and approximate semimajor axis taken from \citet{Janson2014B} unless otherwise noted. \lq \lq?\rq\rq\ in column 4 denotes ambiguity in association membership. YMG / Association ages are taken from \citet{Bell2015} unless otherwise noted.
\\ 
 \hfill

*Sources are partially resolved or previously resolved at small separations indicating rapid orbital periods \citep{Kohler2001, Kohler2002, Daemgen2007}.
 \\
 \hfill
 
 ** J053018 and J201633 YMG membership is highly unlikely following our analysis detailed in sections \ref{s:j053018} and \ref{s:j201633} respectively.
 \\
 \hfill

$^1$Integrated spectral type (Int SpT); \citet{Alonso2015}, $^2$\citet{Janson2007}, $^3$Int SpT; \citet{Riaz2006}, $^4$\citet{Bonnefoy2009}, $^5$Int SpT; \citet{Chauvin2010},
$^6$ Int SpT; \citet{Carpenter2006}, $^7$\citet{Shkolnik2010}, $^8$\citet{Montet2015}, $^{9}$\citet{Konopacky2007},
$^{10}$\citet{Bailey2012}, $^{11}$\citet{Bonavita2016}, $^{12}$\citet{Laf2014}, $^a$\citet{Malo2013}, $^b$\citet{Torres2008}, $^c$\citet{Pecaut2012}
 \\
 \hfill
 
\textbf{References} -- B16; \citet{Bergfors2016}, J14; \citet{Janson2014B}, LM13; \citet{LopezM2013}, LS06; \citet{LopezS2006}, M13; \citet{Malo2013}, M14; \citet{Malo2014}, R15; \citet{Rizzuto2015}, W94; \citet{Walter1994} }\\
\end{table*}

\section{Observations and data reduction}
\label{sec:obs}

All of our RV monitoring was conducted using the Fiberfed Extended Range Optical Spectrograph \citep[FEROS;][]{ Kaufer1999} mounted at the ESO-MPG 2.2 m telescope at La Silla Observatory. The observations used in this survey were taken in service mode under programs 093.A-9006(A) and 094.A-9002(A) between September 2014 and October 2015. A maximum of five spectra were taken per target as an optimal trade-off between initial number of RV data points and required observational time. We also choose to include several archival FEROS observations to bolster the number of spectra for targets in which the ideal sample of five was not obtained, for future orbital analysis. RV measurements for these archival spectra have not been previously published and are included in Table \ref{tab:2} with MJD prior to 56912. A further five spectra of J07285137-3014490 were included (taken from archival and ongoing RV monitoring campaigns) in order to significantly increase the number of available RV measurements, enabling a dedicated study of the system (Rodet et al., in submitted). FEROS is an echelle spectrograph covering the wavelength range $3500-9200$ Angstrom across 39 orders with $R \approx 48 000$. Observations were carried out in \lq \lq object + calibration configuration\rq\rq\ in which one of the two FEROS optical fibres is centered on the star whilst the other simultaneously observes a ThAr+Ne lamp to monitor spectrograph stability. Afternoon calibrations such as bias frames and flat-fields were also taken for data reduction.  

All standard spectroscopic data reduction procedures such as flat-fielding, pixel correction, order extraction and ThAr+Ne lamp wavelength calibration were carried out using the ESO-MIDAS based FEROS Data Reduction System (DRS). The DRS also carries out a re-binning and order merging to produce an optimally reduced continuous spectrum. \citet{Muller2013} note that the automatic barycentric correction the DRS applies is inaccurate. Therefore we reversed the DRS correction and applied an improved correction based on the algorithm of \citet{Stumpff1980}.

In order to derive radial velocity measurements for the sample we cross correlated our observed spectra with synthetic template spectra. Our synthetic spectra were generated from the spectral libraries of \citet{Husser2013}. This library is particularly well suited for our purposes as it has been compiled using the ACES equation of state, which accounts for the formation of molecules at the low temperatures of M-dwarf atmospheres. For each target we used a single synthetic template for cross-correlation. The template was generated using surface gravity, temperature and metalicity input parameters. We adopted a solar metalicity for each target as our sample resides in the solar neighborhood. We adopted temperature and surface gravity values from \citet{Baraffe1998} evolutionary models based on estimated target age and mass. These masses are first order approximations, derived from primary component spectral type for resolved stars, and integrated spectral type for unresolved stars using the spectral type-mass relations of \citet{Kraus2007}. These spectral types are estimated via AstraLux photometry \citet{Janson2012} following the methods of \citet{Daemgen2007}, unless otherwise noted in Table \ref{t:1}.

We computed cross-correlation functions for each of the spectra over multiple wavelength ranges. These ranges correspond to the wavelengths of individual echelle orders that are free from telluric and strong stellar emission lines. $\sim 30$ orders are suitable for cross correlation for each target, although this value decreases for later spectral types due to lower signal to noise data at shorter wavelengths. The edges of each order range were also clipped to avoid edge of chip effects and any inaccuracy in the DRS order merging. The RV was measured across each range by fitting a Gaussian to the CCF, or multiple Gaussians for multiple component spectroscopic binaries. We then derived a final RV measurement for each spectrum by taking a mean of these individual RV measurements (weighted by the goodness of the Gaussian fit to each CCF) and calculating the uncertainty by computing the standard error.

\section{Results and discussion}
\label{sec:res}

\begin{figure}
\centering
\includegraphics[width=8.5cm]{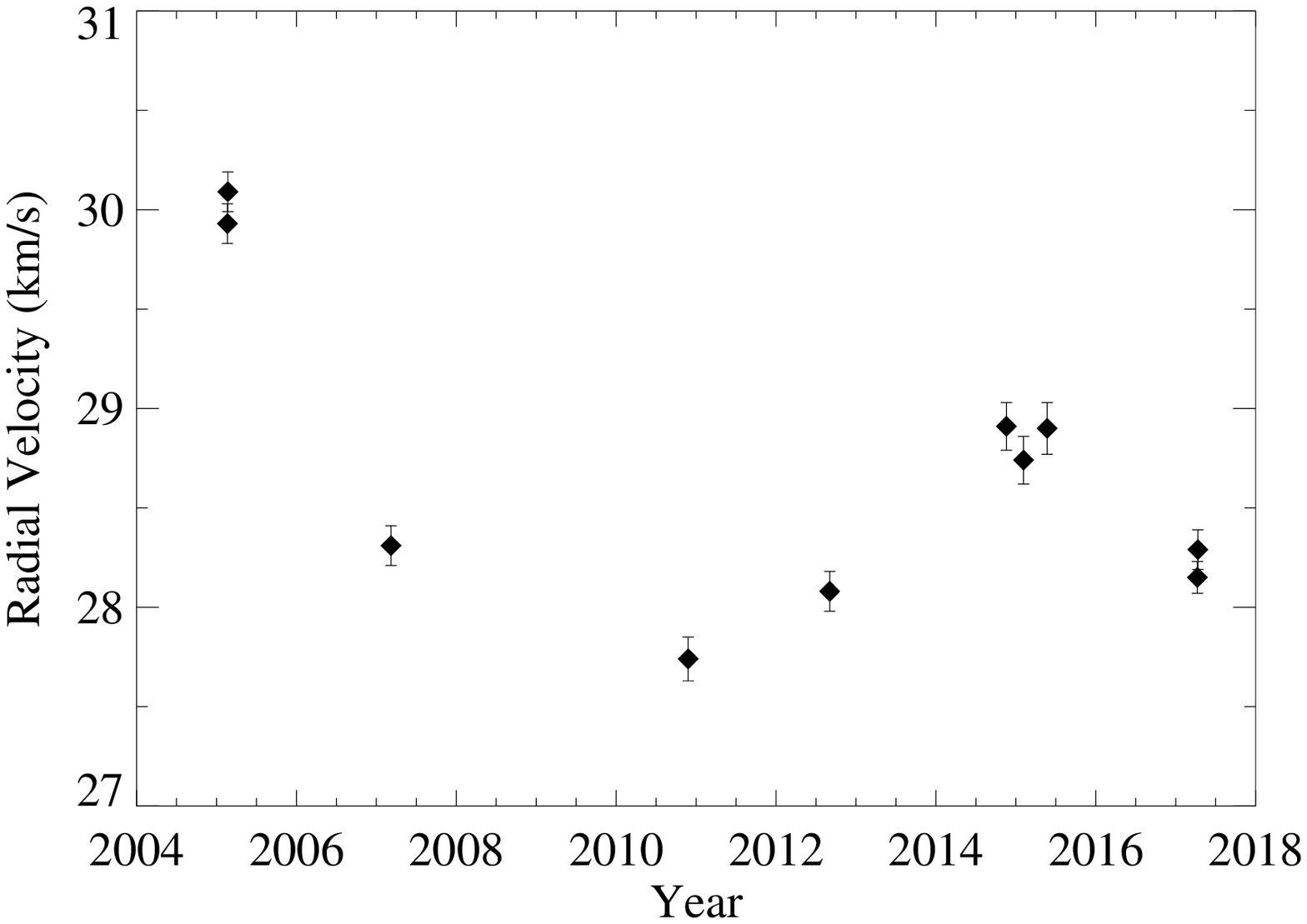}
\caption{ Example RV plot featuring data measured for the single-lined J07285137-3014490 system. RV motion is evident and will be combined with astrometric data in order to constrain the system orbit; Rodet et al. (submitted). The three data points between 2014 and 2016 were taken as part of this survey (programs 093.A-9006(A) and 094.A-9002(A)), the remainder come from archival and ongoing surveys.}
\label{f:rv}
\end{figure}

The results of our RV monitoring survey for our sample of 29 M-dwarf multiples are presented in Table \ref{tab:2}. On average, we derive individual RVs to a precision of $\sim0.2$ kms$^{-1}$, obtaining $\sim3-4$ epochs of measurements per target. An example RV plot is shown in Figure \ref{f:rv}. RV's for the targets J024514, J090758 and J234953 are reported here for the first time, whilst several other targets have RV's reported to sub km/s precision for the first time. There are two three-component spectroscopic multiples present in this sample, one of which is a new discovery, and which is discussed in the section \ref{s:j053018}. In section \ref{s:j201633} we also present the new discovery of a suspected single-lined triple system. Further individual targets are discussed in Appendix A.

\subsection{High-order multiplicity of 2MASS J05301858-5358483} 
\label{s:j053018}

\begin{figure}
\centering
\includegraphics[width=8.5cm]{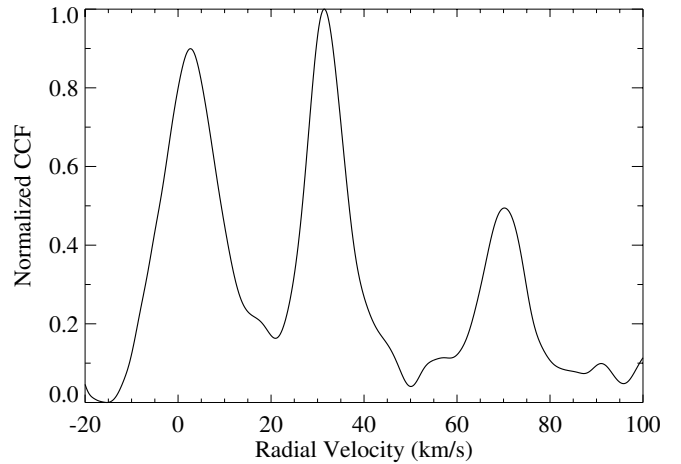}
\caption{ Example CCF plot for the J053018 system. Three individual peaks are clearly distinguishable, and therefore we identify the system as a three-component spectroscopic multiple. Spectroscopic components are identified by tracing CCF peak strength across each epoch. We relate the strongest peak to the A component, occurring at $\sim30\ $ kms$^{-1}$, the $2^{nd}$ strongest peak to the B component, occurring at $\sim0\ $kms$^{-1}$, and the lowest strength peak to C component, occurring at $\sim70\ $kms$^{-1}$. The CCF displayed has been measured for MJD = 55525.295, across the $36^{th}$ spectral order.}
\label{f:ccf_2}
\end{figure}

\begin{figure}
\includegraphics[width=9cm,height=6cm]{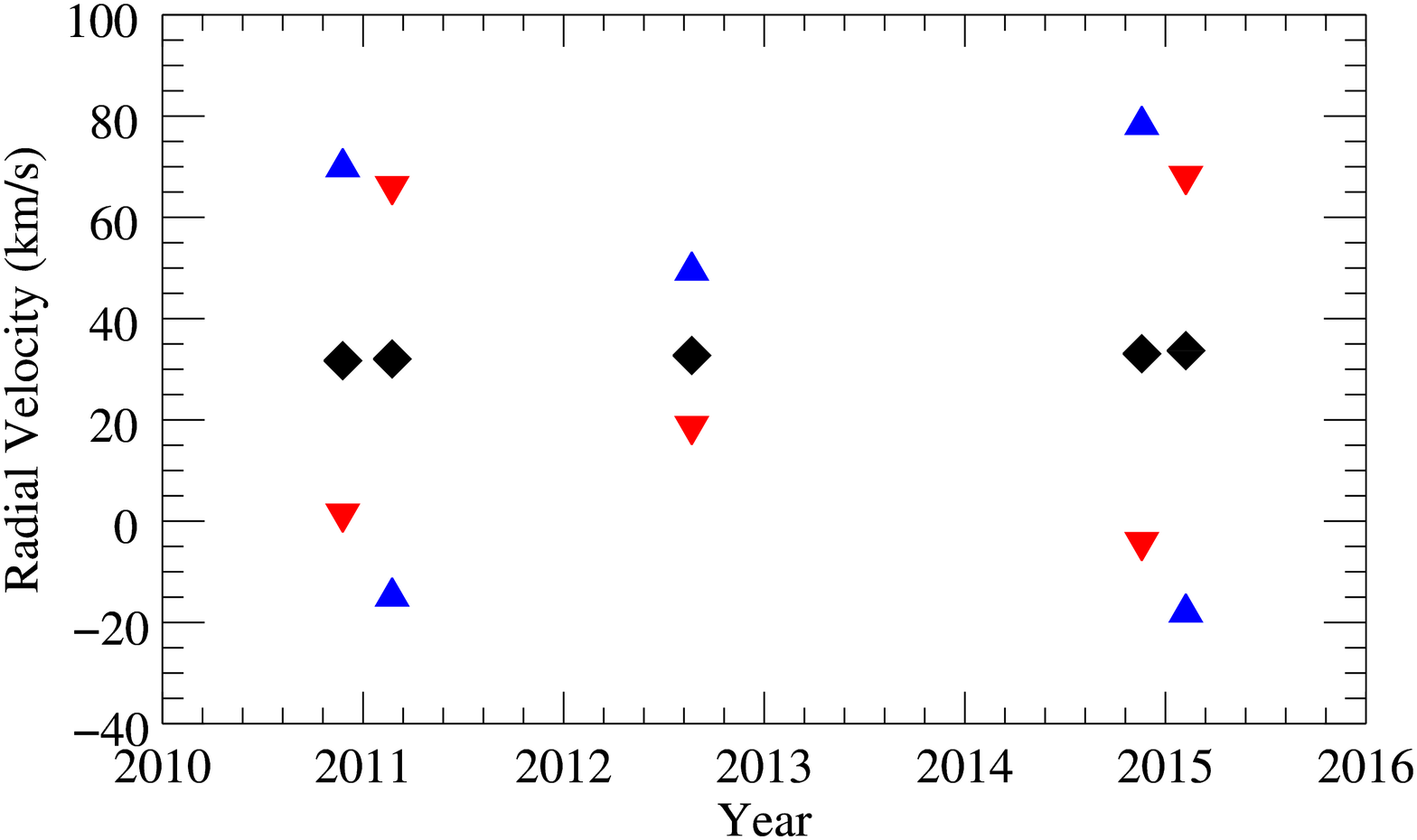}
\caption{ RV data for the J053018 system. Errors are on the order of $\sim0.3$ kms$^{-1}$ and therefore lie within the data point boundaries. Color relates the measured RVs to indiviudal spectroscopic components; black = A component, red = B component, blue= C component, see text for details.}
\label{f:3rv}
\end{figure}

\begin{figure}
\hspace{-1.5cm}
\includegraphics[width=11cm]{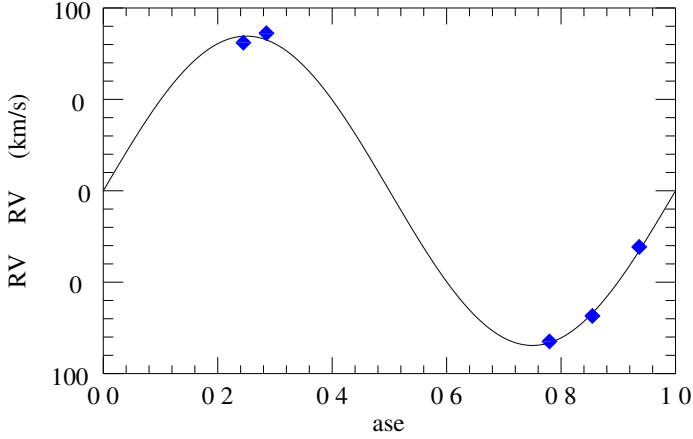}
\caption{Preliminary orbital fit to the J053018 BC system. BC component relative RVs ($RV_C - RV_B$) are plotted in blue as a function of orbital phase. Errors are on the order of $\sim0.4$ kms$^{-1}$ and therefore lie within the data point boundaries. Data is fit by a sine curve with a period of $P_{\rm prel} = 3.4$ days, and a semi-amplitude of $K_{\rm prel} = 84.6$ km/s.  }
\label{f:53018_orbit}
\end{figure}

The system 2MASS J05301858-5358483, hereafter abbreviated as J053018, was previously known to be a M3.0+M4.0+M6.0 resolved triple system. In each of our RV epochs we recover a triple-peaked CCF and therefore identify J053018 as a three-component spectroscopic multiple. An example triple-peaked CCF is shown in Figure \ref{f:ccf_2}. \citet{Janson2012, Janson2014B} derive a separation of $\sim0.2\arcsec$ for the tight M3.0+M4.0 pair, and a separation of $\sim4\arcsec$ for the wide M6.0 companion. As the M6.0 companion falls outside the $2.0\arcsec$ FEROS fibre aperture, the third spectroscopic component is most likely due to an additional unresolved tight companion, making J053018 a quadruple system. As mentioned in Section \ref{sec:intro}, identifying any unresolved components in a multiple system is of critical importance for an accurate isochronal analysis. The RV data for J053018 is plotted in Figure \ref{f:3rv}. We identify which spectroscopic component is responsible for each individual RV measurement by tracing CCF peak strength across each epoch. We relate the strongest peak in each epoch to the \lq \lq A\rq\rq\ component, occurring at $\sim30\ $ kms$^{-1}$ in Figure \ref{f:ccf_2}. The $2^{nd}$ strongest CCF peak we relate to the \lq \lq B\rq\rq\ component and the lowest strength peak to \lq \lq C\rq\rq\ component, occurring at $\sim0\ $kms$^{-1}$ and $\sim70\ $kms$^{-1}$ respectively in Figure \ref{f:ccf_2}. Figure \ref{f:3rv} displays the slowly varying RV motion of the A component, indicating that it is the $\sim0.2\arcsec$ separation resolved component, whilst the B and C components display rapid RV motion, indicative of an unresolved tight spectroscopic binary. Figure \ref{f:3rv} also indicates that the B + C pair are in anti-phase and have gone through $\sim180$ degrees of motion, or some integer multiple of 180 degrees, between the 2010 and 2011 epochs (separated by $\sim90$ days) and the 2014 and 2015 epochs (separated by $\sim80$ days). Therefore, to first order we estimate the longest possible orbital period of the B + C pair to be $\sim170$ days, and possibly factors of several shorter as multiple orbits may have been completed over the $\sim$ 80 to 90 day baseline. Since the sampling of the RV data is very coarse relative to the period of the orbit, it is premature to attempt an exact orbital fitting. However, in order to acquire a tentative overview of which families of orbits would be feasible, we fit simple sine curves to the data, sampling the period from 1 to 170 days. In doing so, we consider the BC pair with the frame of reference fixed on component B, meaning we fit the sine curve to the $RV_C - RV_B$ data points. We find a best fit (minimum quadrature sum of the residuals) period of $P_{\rm prel} = 3.4$ days, and a semi-amplitude of $K_{\rm prel} = 84.6$ km/s. The phase-folded fit is shown in Figure \ref{f:53018_orbit}. We reiterate that this is a preliminary fit and that future RV monitoring with a denser sampling will be needed for an unambiguous orbit determination. However, it is interesting to note that $P_{\rm prel}$ and $K_{\rm prel}$ are mutually consistent: If we assume a BC system mass of 0.3~$M_{\rm sun}$, based on individual spectral type (see following paragraph) and the relations of \citet{Kraus2007}, and a circular orbit, then the expected edge-on RV semi-amplitude for an orbital period of 3.4 days is 96.3 km/s. In projection, this is consistent with $K_{\rm prel} = 84.6$ km/s if the inclination of the BC orbit is $\sim$60 degrees. By contrast, if the orbital period were 10 days, then the expected edge-on RV semi-amplitude would be 65.5 km/s. This is too low to match the existing data points for a circular orbit. For longer periods, this discrepancy becomes increasingly pronounced. Hence a period in the $<$10~d range is favored by this argument.

\begin{figure}
\includegraphics[width=8cm]{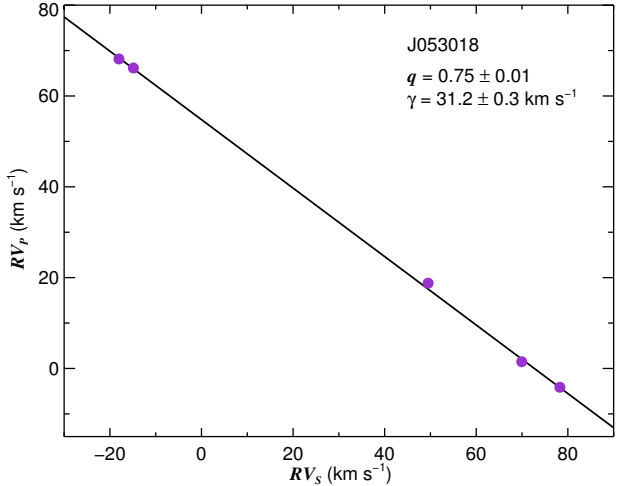}
\caption{J053018 B and C component RVs, denoted as RV$_P$ and RV$_S$ respectively. The negative gradient of the best-fit line is equal to the
mass ratio of the system ($q = 0.75 \pm 0.01$), and we measure a systemic RV ($\gamma$) of
$31.2 \pm 0.3\ $kms$^{-1}$.}
\label{f:mr}
\end{figure}

Regardless of the specific orbit, as we can distinguish the B and C component RVs, we can measure the mass ratio and systemic RV of the pair following the methods of \citet{Wilson1941}. The component RVs are plotted against each other in Figure \ref{f:mr} where the mass ratio of the pair is given by the negative of the gradient of the line. We measure a mass ratio of $0.75 \pm 0.01$ and a systemic RV of $31.2 \pm 0.3\ $kms$^{-1}$ for the pair. This systemic RV is consistent with the $31.3 \pm 0.2\ $kms$^{-1}$ RV measured for the M3.0 primary by \citet{Malo2014}. This \citet{Malo2014} measurement  is also consistent with the RV we measure for the slowly varying A component. We therefore identify the A component as the M3.0 primary and the M4.0 companion as the unresolved BC pair. We suspect the BC components are comprised of an M4.0 + M5.0 pair. This is based on relating the mass ratio of the pair, and the relative strengths of the CCF peaks, to the unresolved M4.0 spectral subtype. 
 
\begin{figure*}
 \centering
\includegraphics[width=16cm]{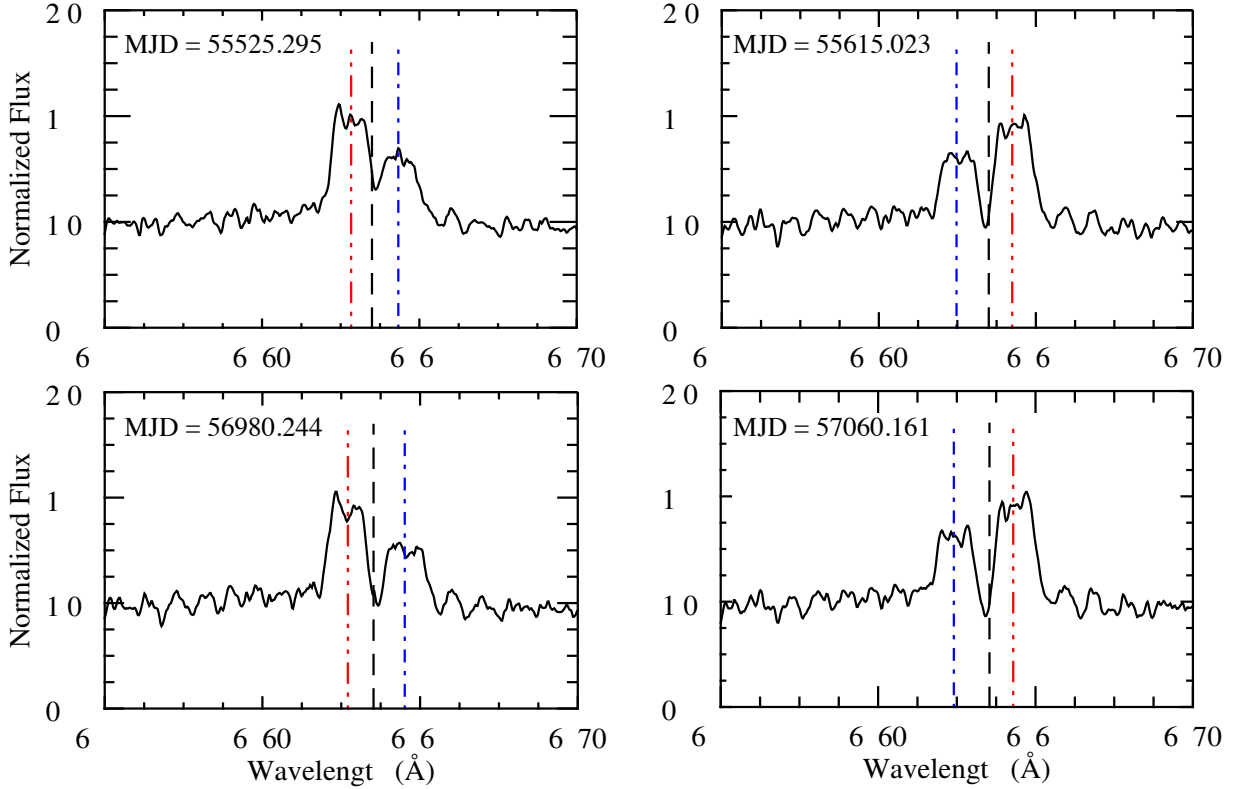}
\caption{Four epochs of J053018 FEROS spectra covering H$\alpha$ emission wavelengths. The dashed lines highlight the predicted positions of the H$\alpha$ emission line given the measured RVs of the A (black line), B (red line) and  C (blue line)  components. In each epoch, prominent emission lines are visible at the predicted H$\alpha$ position given the measured RVs of the BC pair, whilst no emission is visible at the predicted H$\alpha$ position given the measured RV of the A component. }
\label{f:ha}
\end{figure*}

J053018 was previously thought to be a high probability member of the AB Dor YMG based on its kinematics with young age supported by significant X-ray emission \citep[e.g.,][]{Lepine2011, Janson2012}. Using proper motions and their RV measurement for the M3 primary, \citet{Malo2014} derive an AB Dor membership probability of 97.7 \% using Bayesian Analysis for Nearby Young AssociatioNs \citep[BANYAN,][]{Malo2013}, a statistical tool tracing YMG membership based on Galactic position and space velocity.  As no trigonometric distance measurement exists for J053018, the BANYAN tool marginalizes this parameter in the YMG probability determination and generates the most probable statistical distance, $3 \pm\ 1$ pc, assuming membership to the AB Dor YMG is bona fide. We reevaluate the YMG membership using the BANYAN $\Sigma$ tool \citep{Gagne2018} and the systemic RV we measure for the BC pair. We find an AB Dor membership probability of $> 95$\% and a statistical distance of $2.5 \pm 0.8$ pc in agreement with the \citet{Malo2014} findings. However, whilst no trigonometric distance exists for J053018, \citet{Janson2012} derive a photometric distance of $23 \pm 9$ pc for the system. Unlike the statistical distance, this measurement is independent of any YMG membership assumption. At  $23 \pm 9$ pc J053018 is highly unlikely to be an AB Dor member given its other kinematics and indeed, the BANYAN $\Sigma$ tool returns a 0\% membership probability. We suspect that this system is instead an old field interloper that appears young and X-ray active because the rotational velocities of the tight BC pair are high, due to spin-orbit locking. Enhanced chromospheric and coronal emission due to spin-orbit locking has previously been observed for the synchronized binary BF Lyn with an orbital period of $\sim3.8$ days \citep[e.g.,][]{Montes2000, Maldonado2010}. This argument is supported by a visual inspection of our FEROS spectra over H$\alpha$ emission wavelengths. Figure \ref{f:ha} shows the H$\alpha$ emission region of several epochs of spectra. In each epoch prominent emission lines are visible at the predicted H$\alpha$ position given the measured RVs of the B and C components. However, no significant emission is visible at the predicted H$\alpha$ position given the measured RV of the A component. This strongly suggests that the A component is no longer active because it is old, whilst the B and C components are active due to tidal spin-up. This argument is further reinforced by an inspection of the CCF peaks. As seen in Figure \ref{f:ccf_2}, the B and C component peaks are significantly, $\sim30$ \%, broader than the A component peak. This is true across each epoch and usable order. Whilst this may be due to a mis-match with the cross correlation template, this suggests that the BC pair have larger rotational velocities than the A component, supporting the theory that the pair has been spun-up. The  $P_{\rm prel} = 3.4$ day period is sufficiently short to allow the rotation of the components to be tidally locked to their orbit, and thus to be spun-up sufficiently to produce X-ray emission that matches the activity levels of a young ($< 100$ Myr) population \citep[e.g.,][]{Herbst2007}. \\

\subsection{The case of 2MASS J20163382-0711456} 
\label{s:j201633}

\begin{figure}
\includegraphics[width=9cm]{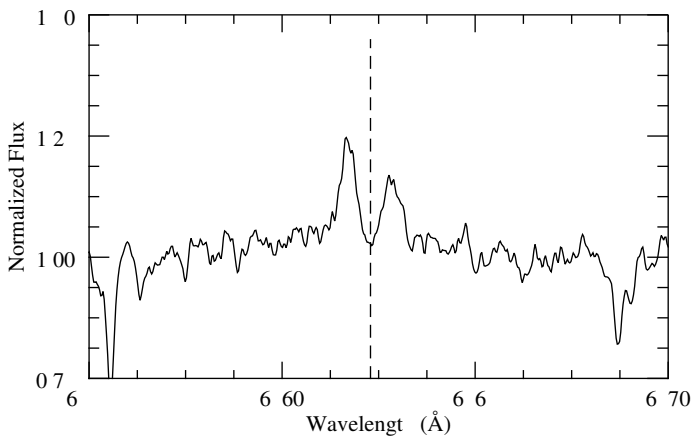}
\caption{J201633 FEROS spectra covering H$\alpha$ emission wavelengths. The dashed line highlights the predicted position of the H$\alpha$ emission line given the measured RV of the M0.0 primary. No emission is visible at this position. However, emission lines are observed shifted to the blue and red of this position. This suggests the M0.0 primary is no longer active because the system is old whilst the M2.0 secondary is an unresolved synchronized pair which is active due to tidal spin-up, resulting in  H$\alpha$ emission shifted according to the RV of the individual components.  }
\label{f:ha_2}
\end{figure}

The system 2MASS J20163382-0711456, hereafter abbreviated as J201633, was previously known to be a M0.0+M2.0 resolved binary and a probable member of the Argus YMG. With a  separation of $\sim 0.1\arcsec - 0.2\arcsec$, both components fall within the FEROS aperture. In each of our RV epochs we recover a single-peaked CCF and therefore identify J201633 as a single-lined binary. However, similar to the case of J053018, visual inspection of our FEROS spectra over H$\alpha$ emission wavelengths suggests one of the components hosts an unresolved synchronized companion. Figure \ref{f:ha_2} shows the H$\alpha$ emission region of a single epoch, which is representative of the series of observations. Again, no significant emission is visible at the predicted H$\alpha$ position given the measured RV of the single lined system, however, emission lines are observed shifted to the blue and red of this position. We suspect the M0.0 primary is no longer active because the system is old whilst the M2.0 secondary is an unresolved synchronized pair which is active due to tidal spin-up, suggesting the system is an old field interloper. Due to the low signal to noise and similar heights of the resolved H$\alpha$ peaks, we are unable to trace peak height across each epoch and relate specific H$\alpha$ peaks to individual binary components. 

Multiple studies \citep[e.g.,][]{Malo2013, Malo2014, Janson2014B} have consistently identified J201633 as a probable Argus member using the BANYAN I and II tools, proper motions from the UCAC3 catalog \citep[R.A. = $84.6 \pm 8.1$ mas/yr, decl. = $-0.6 \pm 23.6$ mas/yr,][]{Zacharias2009} and both marginalized (i.e., omitted) and photometric distances. We reevaluate the YMG membership using the BANYAN $\Sigma$ tool and refined UCAC4 catalog proper motions \citep[R.A. = $71.5 \pm 3.9$ mas/yr, decl. = $13.8 \pm 3.5$ mas/yr,][]{Zacharias2012} and find a 99.9\% probability that J201633 is a field star, given both marginalized and photometric \citep[d = $50 \pm 18.5 $ pc][]{Janson2014B} distance input parameters. However, the Argus group has been removed entirely from the models of BANYAN $\Sigma$, as it has been demonstrated to be composed of non-coeval stars (see section \ref{sec:samp}). Given the BANYAN $\Sigma$ field star probability and lack of H$\alpha$ emission from the primary, it is highly likely J201633 is an old field interloper previously associated with the now doubtful Argus association. BANYAN $\Sigma$ also returns a $\sim99.9$\% field star probability for the remaining targets in the sample previously associated with Argus (see Table \ref{tab:2}). However, each of these targets displays strong H$\alpha$ emission due to the primary component and is highly unlikely to host a synchronizing companion around the primary (see section \ref{s:synch}). Therefore whilst these targets are young ($< 100$ Myr), it is unclear if they are associated to any kinematic group. Unlike J053018, we do not detect any spectroscopic lines beyond H$\alpha$ for the unresolved pair, and therefore cannot derive any individual RV measurements or generate an orbital fit. However, similar to J053018, we suspect the unresolved pair has a significantly short period, on the order of $1 - 10$ days, in order for tidal synchronization to enhance X-ray emission that matches the activity levels of a young ($< 100$ Myr) population \citep[e.g.,][]{Herbst2007}.

\subsection{Limits on tidal synchronization}
\label{s:synch}

As the cases of J053018 and J201633 show, tidally synchronized pairs can mimick signatures of youth in low-mass systems, and can therefore critically bias isochronal analyses unless identified and discarded. Hence, we have scanned our sample for additional pairs of this nature, but only J053018, J201633 and the previously known J042442 system  (with upper limit of 1.9 days on the orbital period of the BC pair, see Appendix A) show any such indications. In general, such systems should be easily identified, because at the necessary periods of $<$10~d for tidal synchronization to occur over $\sim$100~Myr timescales \citep[see e.g.,][]{Meibom2006}, they can be expected to have velocity semi-amplitudes of $>$30~km/s, similar to the J053018 system. However, there are two effects that can hide an otherwise detectable pair: (i) The system inclination may be very close to face-on, leading to a very small fraction of the total velocity to be projected into the measurable radial component. (ii) Repeated RV measurements may accidentally sample the orbit in phases with similar instantaneous RV values. Both of these are low-probability effects, and the latter is particularly improbable if the RV has been sampled multiple times. Nonetheless, in a large enough sample, they should be expected to occur to some level, and it is therefore desirable to quantify to which extent they can be excluded for the targets in our sample without any detected signature. For this purpose, we have performed a series of Monte Carlo simulations for each individual target, as described in the following. 

The total velocity semi-amplitude $K_{\rm tot}$ of a binary pair is a function of mass $M_{\rm s}$ of each individual known star that is hypothesized to host a synchronizing companion (hereafter named a `synchronizer'), the mass ratio $q$ of the pair, and their eccentricity $e$ and orbital period $P$. Whilst J042442 (see Appendix A), J053018 and J201633 host synchronizing companions around their secondaries, the remainder of the sample are identified as single-lined spectroscopic binaries that display H$\alpha$ emission at the expected position given the measured RVs. Therefore if a synchronizing companion mimicking signatures of youth is present in these systems, it must orbit the primary component. Therefore, we only evaluate the presence of an unresolved synchronizer around this component in each system. Hence, $K_{\rm tot}$ is solely a function of the primary mass $M_{\rm s}$, $q$, $e$ and $P$. Since we are interested only in tidally synchronized pairs in this analysis, the orbital periods will be limited to $P < 10$~d as mentioned above. Setting $P = 10$~d will thus provide a lower limit on $K_{\rm tot}$. In such short-period systems, eccentricities are known to be very low \citep[e.g.,][]{Raghavan2010}, which is a natural consequence of the same tidal forcing that causes the period synchronization. Hence, we set $e = 0$ in our simulations. The mass ratio distribution of close-in and low-mass systems is strongly peaked toward $q = 1$, and do not show $q < 0.5$ stellar systems even in surveys that are complete below this value \citep[e.g.,][]{Reid1997,Delfosse2004}. Similarly to the $P$ case, we thus set $q = 0.5$ to acquire a lower limit on $K_{\rm tot}$. The $M_{\rm s}$ masses are assigned individually for the targets in our sample, based on their spectral types following the same relation as in \citet{Janson2012}. 

For each system, we then perform $10^5$ simulations in which $K_{\rm tot}$ is projected into a radial component assuming random orientations of the orbit. Since our observations are taken at irregular intervals that are generally much longer than 10~d, we can assume that such short orbits are effectively randomly sampled. We thus divide the velocity curve into a fine grid of small temporal segments, and randomly select $N_{\rm obs}$ segments for each of the $10^5$ randomly projected orbits, where $N_{\rm obs}$ is the number of observations available for the target in question. We then evaluate the RV difference between adjacent pairs in the simulated ($\delta K_{\rm sim}$) and real ($\delta K_{\rm obs}$) observations, and choose the maximum among the $N_{\rm obs} - 1$ pairs in both cases. If $\max{\delta K_{\rm sim}} > \max{\delta K_{\rm obs}}$, it is considered that the hypothesized synchronizer should have been detected in this random instance, while the opposite is true if $\max{\delta K_{\rm sim}} < \max{\delta K_{\rm obs}}$. The fraction among the $10^5$ simulations for a given target in which the synchronizer should have been detected is denoted $f_{\rm c}$. If $f_{\rm c}$ is close to 1, it means that there is virtually no way to hide a synchronizer, and so the hypothesis of its existence can be discarded. If $f_{\rm c}$ is small, there is not yet enough data to conclude whether a synchronizer might exist. The motivation for using adjacent pairs of observations as opposed to, for example, the minimum and maximum of the full set of observations is to mitigate the impact of the slower gradual RV trend that arises from the already known wider companions in the system.

All calculated $f_{\rm c}$ values are shown in Table \ref{tab:2}. For the two systems J033235 and J052844, only one epoch of observations exists, and for the J042442, J053018 and J201633 systems a close companion has already been identified, so for those systems no meaningful $f_{\rm c}$ can be calculated. However, to test our methodology we compute $f_{\rm c}$ for the J053018 B component, as if we had not identified and measured RVs for the C component. Under these circumstances $f_{\rm c}$ = 0.0, meaning there is no single case in which a simulated 10 day companion induces a larger RV variability than that measured for J053018  B.  This is consistent, and therefore provides validation to our analysis, with a 10 day period corresponding to the minimal $K_{\rm tot}$ that could be caused by a synchronizer and the 3.4 day orbital period we estimate for the BC pair. For the other systems, $f_{\rm c}$ is generally very high -- the lowest value is 0.799 for J061610, which still implies that it is unlikely that any synchronizer could hide from view, if such an object exists in the first place. The highest values are $>$0.999 for 5 systems, such that the synchronizer hypothesis can arguably be discarded entirely. This analysis implies that youth is the most probable explanation for high chromospheric activity in the vast majority of our sample, as is often supported also through kinematic YMG analysis. For J024514, there is point-to-point RV scatter significantly exceeding the formal errors, which could imply that a close-in companion exists. However, in this case the $f_{\rm c}$ of 0.964 implies that any such companion has a lower RV amplitude than would be expected from a synchronizer. In other words, while an additional unseen companion could exist in that system, it could not be responsible for the high activity level of the system. However, whilst the conclusion of youth remains probable in such a case, the presence of a non-synchronizing, unseen companion could lead to an incorrect isochronal age estimate or model calibration if the unresolved pair is treated as a single star. We estimate the frequency of such companions in our sample. The maximum separation limit of an unresolved companion is $\sim2.2$ AU, given the approximate resolving power of AstraLux, $0.1\arcsec$, and the typical target distance, $\sim22$ pc. The minimum separation limit of a non-synchronizing companion is $\sim0.05$ AU, given P = 10 days (as periods below this limit are required for tidal synchronization), a target mass of 0.245 $M_{\sun}$ (from \citet{Kraus2007} for a typical survey target, SpT = M3.5) and q in the range 0.5 - 1.0. \citet{Fischer1992} give an indication of M-dwarf multiplicity over this approximate parameter space as they find 4 companions between 0.04 and 4.0 AU among 62 targets with a detection probability of 86\%. Taken at face value this would give a multiplicity frequency of 7.5\% in this range. Therefore, we adopt 7.5\% as an upper frequency limit on 0.05-2.2 AU companions, as the derived \citep{Fischer1992} frequency covers a broader separation range.

The value of $f_{\rm c}$ depends only weakly on the mass $M_{\rm s}$, but depends quite strongly on the number of data points acquired; with only two data points it is relatively easy to miss a high RV variability if (for example) two epochs are accidentally separated by an integer multiple of the periodicity, but with a larger number of irregularly spaced observations, this becomes exceedingly unlikely. However, high $f_{\rm c}$ values ($> 0.99$) in the cases of J090758 and J205314 are derived as the RV variability has so little scatter that it is highly unlikely to catch a synchronizer with such minimal $\delta K_{\rm obs}$, even over just two randomly sampled data points.

\section{Summary and conclusions}
\label{sec:sum}

In this paper we have presented the results of an RV monitoring survey of a sample of 29 young M-dwarf multiples, many of which are high probability YMG members. These targets provide excellent laboratories for a range of astrophysical investigations and are prime candidates for isochronal analysis that can be used to calibrate low-mass stellar models and date individual systems. This dating can then potentially be applied to the full YMG of which they are members, for improved age constraints on the full population of stars. Our sample has been monitored through various astrometric monitoring campaigns, primarily our AstraLux multiplicity surveys, with the aim of deriving full orbital parameters and dynamical masses for the binary components. These model independent masses are essential for a robust isochronal analysis. Our RV measurements reported here complement the imaging data, allowing enhanced orbital determinations and precise dynamical masses to be derived in a shorter timeframe than possible with astrometric monitoring alone. Furthermore, as illustrated by the case of J053018, RV allows us to identify binary sub-pairs that are close enough to tidally synchronize, which increases their level of chromospheric activity. If such pairs go undetected, the age of the system can be drastically misestimated, which would invalidate any isochronal analysis. We have shown that the majority of our sample exhibits short-term RV variability on a level that is far lower than would be expected if they hosted $P < 10$~d stellar companions, even when accounting for the possiblity of unfortunate orbit projections or non-optimal orbit sampling. Thus, nearly all of our targets are best interpreted as genuinly young binaries with spatially resolvable orbits of a few years to decades, confirming their suitability for isochronal analyses. In future work, detailed orbital analysis and complementary spectral analysis of J043737 and J072851 (Rodet et al., submitted) will be presented.

\begin{acknowledgements}
S.D. acknowledges support from the Northern Ireland Department of Education and Learning. M.J. gratefully acknowledges funding from the Knut and Alice Wallenberg Foundation. C.A.W. acknowledges support by STFC grant ST/P000312/1. This work is based on observations made with FEROS on the ESO-MPG 2.2 m telescope at La Silla Observatory. This study made use of the CDS services SIMBAD and VizieR, as well as the SAO/NASA ADS service.
\end{acknowledgements}

\onecolumn


{\small
\begin{longtable}{llllll}
\caption{Radial Velocity Measurments.}\\
\hline\hline
		2MASS ID &  Epoch & RV &  RV Error & $f_{\rm c}$ & Notes \\
		& (MJD) & (kms$^{-1}$) & (kms$^{-1}$) & (Sect. \ref{s:synch}) &  \\
\hline
\endfirsthead
\caption{continued.}\\
\hline\hline
		2MASS ID &  Epoch & RV &  RV Error & $f_{\rm c}$ & Notes \\
		& (MJD) & (kms$^{-1}$) & (kms$^{-1}$) & (Sect. \ref{s:synch}) &  \\
\hline
\endhead
\hline
\endfoot
\hline
\multicolumn{6}{l}{\footnotesize{\textbf{NOTE} -- SB1 and SB3 denote a singled-lined binary and a three-component spectroscopic multiple}}\\
\multicolumn{6}{l}{\footnotesize{respectively. For SB3 system J042442 we note the RV measured for the central CCF peak, and peaks}}\\
\multicolumn{6}{l}{\footnotesize{toward the blue and red with respect to the central peak, as individual components cannot be}}\\
\multicolumn{6}{l}{\footnotesize{distinguished between epochs. For SB3 system J053018 we note the RVs measured for the identified}}\\
\multicolumn{6}{l}{\footnotesize{ABC components, see text for details.}}\\
\hline
\endlastfoot

J01112542+1526214	&	56915.351	&	3.14	&	0.24	&	0.915	&	SB1	 \\
	&	56978.061	&	3.44	&	0.22	&		&		 \\
\hline
J02451431-4344102	&	56161.302	&	31.59	&	0.14	&	0.964	&	SB1	 \\
	&	56912.268	&	30.74	&	0.18	&		&		 \\
	&	56979.144	&	31.23	&	0.30	&		&		 \\
	&	57058.048	&	31.18	&	0.21	&		&		 \\
	&	57070.103	&	27.99	&	0.21	&		&		 \\
\hline
J02490228-1029220	&	55613.082	&	16.40	&	0.16	&	0.995	&	SB1	 \\
	&	56161.326	&	16.85	&	0.13	&		&		 \\
	&	56978.203	&	16.80	&	0.12	&		&		 \\
\hline
J03323578+2843554	&	56979.165	&	10.73	&	0.20	&	---	&	SB1	 \\
\hline
J04244260-0647313	&	55790.421	&	7.81	&	0.46	&	---	&	SB3, Central Peak	 \\
	&	55790.421	&	-33.58	&	0.37	&		&	SB3, Blue Peak	 \\
	&	55790.421	&	57.52	&	0.49	&		&	SB3, Red Peak	 \\
	&	56912.297	&	-9.36	&	0.37	&		&	SB3, Central Peak	 \\
	&	56912.297	&	-48.67	&	0.48	&		&	SB3, Blue Peak	 \\
	&	56912.297	&	45.70	&	0.44	&		&	SB3, Red Peak	 \\
	&	56979.202	&	3.01	&	0.29	&		&	SB3, Central Peak	 \\
	&	56979.202	&	-50.21	&	0.71	&		&	SB3, Blue Peak	 \\
	&	56979.202	&	37.21	&	0.23	&		&	SB3, Red Peak	 \\
	&	57061.120	&	8.85	&	0.36	&		&	SB3, Central Peak	 \\
	&	57061.120	&	-13.54	&	0.26	&		&	SB3, Blue Peak	 \\
	&	57061.120	&	22.91	&	0.37	&		&	SB3, Red Peak	 \\
\hline
J04373746-0229282	&	53707.288	&	20.40	&	0.09	&	0.996	&	SB1	 \\
	&	55203.146	&	24.25	&	0.09	&		&		 \\
	&	56912.331	&	23.04	&	0.09	&		&		 \\
	&	56979.236	&	22.89	&	0.08	&		&		 \\
	&	57059.095	&	22.89	&	0.08	&		&		 \\
	&	57290.319	&	22.21	&	0.08	&		&		 \\
	&	57291.257	&	22.46	&	0.10	&		&		 \\
\hline
J04595855-0333123	&	56912.344	&	43.17	&	0.21	&	$>$0.999	&	SB1	 \\
	&	56980.125	&	43.03	&	0.17	&		&		 \\
	&	57060.127	&	43.00	&	0.19	&		&		 \\
	&	57291.272	&	42.80	&	0.19	&		&		 \\
\hline
J05284446-6526463	&	56701.124	&	30.08	&	0.31	&	---	&	SB1	 \\
\hline
J05301858-5358483	&	55525.295	&	31.69	&	0.22	&	---	&	SB3, A component	 \\
	&	55525.295	&	1.48	&	0.24	&		&	SB3, B component	 \\
	&	55525.295	&	69.93	&	0.28	&		&	SB3, C component	 \\
	&	55615.023	&	32.03	&	0.29	&		&	SB3, A component	 \\
	&	55615.023	&	66.16	&	0.36	&		&	SB3, B component	 \\
	&	55615.023	&	-14.87	&	0.46	&		&	SB3, C component	 \\
	&	56161.374	&	32.72	&	0.15	&		&	SB3, A component	 \\
	&	56161.374	&	18.78	&	0.37	&		&	SB3, B component	 \\
	&	56161.374	&	49.49	&	0.35	&		&	SB3, C component	 \\
	&	56980.244	&	33.07	&	0.23	&		&	SB3, A component	 \\
	&	56980.244	&	-4.14	&	0.21	&		&	SB3, B component	 \\
	&	56980.244	&	78.27	&	0.32	&		&	SB3, C component	 \\
	&	57060.161	&	33.67	&	0.22	&		&	SB3, A component	 \\
	&	57060.161	&	68.15	&	0.26	&		&	SB3, B component	 \\
	&	57060.161	&	-18.03	&	0.27	&		&	SB3, C component	 \\
\hline
J05320450-0305291	&	55526.282	&	24.26	&	0.13	&	$>$0.999	&	SB1	 \\
	&	55615.041	&	24.24	&	0.12	&		&		 \\
	&	56164.407	&	24.80	&	0.14	&		&		 \\
	&	56980.258	&	24.82	&	0.14	&		&		 \\
	&	57059.134	&	25.23	&	0.13	&		&		 \\
\hline
J06112997-7213388	&	56980.288	&	19.33	&	0.29	&	0.943	&	SB1	 \\
	&	57058.223	&	19.54	&	0.26	&		&		 \\
\hline
J06134539-2352077	&	55522.312	&	21.28	&	0.21	&	$>$0.999	&	SB1	 \\
	&	56168.403	&	22.07	&	0.25	&		&		 \\
	&	56700.142	&	22.90	&	0.19	&		&		 \\
	&	56980.335	&	22.91	&	0.23	&		&		 \\
	&	57058.209	&	23.11	&	0.23	&		&		 \\
\hline
J06161032-1320422	&	56983.199	&	31.56	&	0.46	&	0.799	&	SB1	 \\
	&	57058.174	&	28.16	&	0.54	&		&		 \\
	&	57291.338	&	31.20	&	0.44	&		&		 \\
\hline
J07285137-3014490	& 53421.159 & 29.93 & 0.10 & $>$0.999 & SB1 \\
         &     53423.153        &      30.09        &      0.10  &               &      \\
         &     54168.043        &      28.31        &      0.10  &               &      \\
         &	55526.355	&	27.74	&	0.11	&		&		 \\
	&	56173.407	&	28.08	&	0.10	&		&		 \\
	&	56980.349	&	28.91	&	0.12	&		&		 \\
	&	57058.295	&	28.74	&	0.12	&		&		 \\
	&	57166.001	&	28.90	&	0.13	&		&		 \\
	&      57853.031       &      28.15        &       0.08  &              &     \\
	&      57855.144      &      28.29        &       0.10   &              &      \\
\hline
J08475676-7854532	&	57058.309	&	18.75	&	0.15	&	0.943	&	SB1	 \\
	&	57118.129	&	18.51	&	0.15	&		&		 \\
\hline
J09075823+2154111	&	57059.178	&	-2.66	&	0.09	&	0.994	&	SB1	 \\
	&	57060.175	&	-2.68	&	0.10	&		&	SB1	 \\
\hline
J09164398-2447428	&	56984.343	&	21.21	&	0.12	&	0.997	&	SB1	 \\
	&	57059.297	&	20.43	&	0.15	&		&		 \\
	&	57060.209	&	20.54	&	0.15	&		&		 \\
	&	57166.015	&	19.66	&	0.16	&		&		 \\
\hline
J10140807-7636327	&	57058.344	&	17.16	&	0.22	&	0.972	&	SB1	 \\
	&	57118.165	&	17.25	&	0.22	&		&	SB1	 \\
\hline
J10172689-5354265	&	57059.310	&	13.05	&	0.23	&	0.998	&	SB1	 \\
	&	57059.330	&	12.96	&	0.25	&		&		 \\
	&	57060.222	&	13.19	&	0.22	&		&		 \\
	&	57166.032	&	12.75	&	0.21	&		&		 \\
\hline
J11315526-3436272	&	56708.278	&	12.13	&	0.24	&	0.997	&	SB1	 \\
	&	56816.146	&	11.90	&	0.20	&		&		 \\
	&	57062.297	&	11.86	&	0.26	&		&		 \\
\hline
J12072738-3247002	&	56816.160	&	9.33	&	0.22	&	$>$0.999	&	SB1	 \\
	&	57060.364	&	9.18	&	0.15	&		&		 \\
	&	57062.285	&	9.04	&	0.15	&		&		 \\
\hline
J12202177-7407393	&	56809.106	&	14.71	&	0.17	&	0.990	&	SB1	 \\
	&	57060.294	&	15.28	&	0.17	&		&		 \\
	&	57118.201	&	15.02	&	0.17	&		&		 \\
\hline
J13493313-6818291	&	56809.141	&	14.75	&	0.19	&	0.993	&	SB1	 \\
	&	57060.330	&	13.84	&	0.15	&		&		 \\
	&	57062.310	&	13.81	&	0.18	&		&		 \\
	&	57118.262	&	15.07	&	0.18	&		&		 \\
\hline
J15573430-2321123	&	56809.190	&	-5.49	&	0.10	&	0.970	&	SB1	 \\
	&	57119.295	&	-5.62	&	0.16	&		&		 \\
\hline
J20163382-0711456	&	56912.124	&	-23.31	&	0.06	&	---	&	SB1	 \\
	&	56982.023	&	-23.10	&	0.06	&		&		 \\
	&	57166.263	&	-23.40	&	0.06	&		&		 \\
	&	57290.074	&	-23.93	&	0.06	&		&		 \\
	&	57291.068	&	-23.53	&	0.06	&		&		 \\
\hline
J20531465-0221218	&	56912.138	&	-39.57	&	0.13	&	0.996	&	SB1	 \\
	&	56979.057	&	-39.57	&	0.16	&		&		 \\
\hline
J23172807+1936469	&	56912.207	&	-0.06	&	0.14	&	0.861	&	SB1	 \\
	&	56979.091	&	-0.79	&	0.13	&		&		 \\
\hline
J23205766-0147373	&	56809.310	&	-5.99	&	0.17	&	0.984	&	SB1	 \\
	&	56912.173	&	-6.43	&	0.17	&		&		 \\
	&	56980.052	&	-5.93	&	0.17	&		&		 \\
\hline
J23495365+2427493	&	56912.233	&	-8.92	&	0.17	&	0.976	&	SB1	 \\
	&	56979.104	&	-9.13	&	0.14	&		&		 \\
	&	57291.090	&	-8.37	&	0.24	&		&		 \\

\label{tab:2}
\end{longtable}
}

\begin{appendix}
\section{Individual target discussion}

\begin{figure}
\centering
\includegraphics[width=8.5cm]{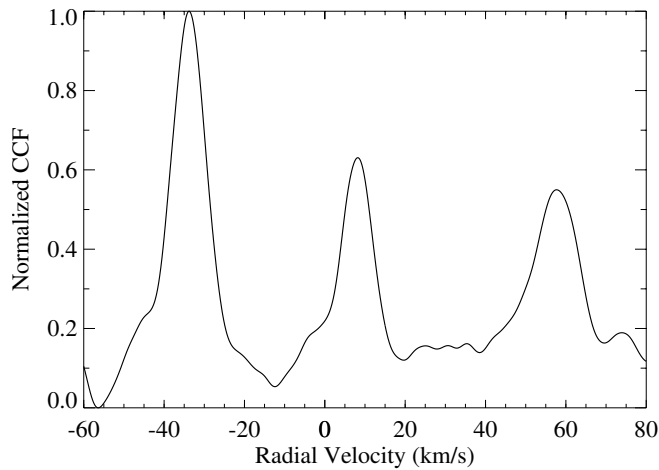}
\caption{ Example CCF plot for the J042442 system. Three individual peaks are clearly distinguishable, and therefore we identify the system as a three-component spectroscopic multiple. The CCF displayed has been measured for MJD = 55790.421, across the $35^{th}$ spectral order.}
\label{f:ccf_1}
\end{figure}

\begin{figure}
 \hspace{3cm}
\includegraphics[width=12cm]{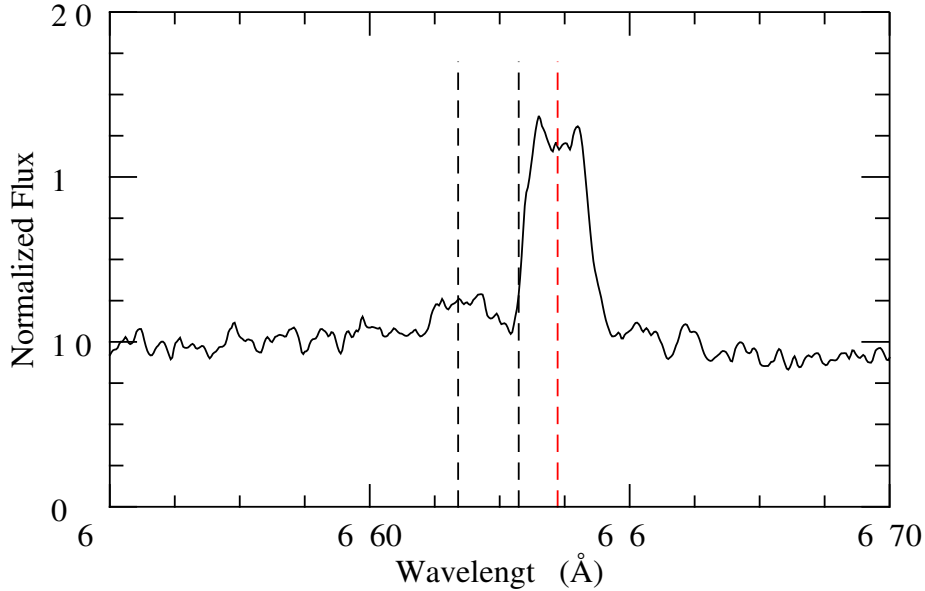}
\caption{J042442 FEROS spectra covering H$\alpha$ emission wavelengths. The dashed line highlights the predicted position of the H$\alpha$ emission line given the measured RVs of the M5.5 and M5.7 BC pair (black lines - components indistinguishable) and the M4.5 primary (red line). Unlike J053018 and J201633, J042442 displays strong H$\alpha$ emission at the predicted position given the measured RV of the primary.  H$\alpha$ emission is tentatively detected given the measured RV of one of the BC components and undetected given the other. This is a reasonable given the brightness contrast between the primary and the BC pair and the apparent RV separations.}
\label{f:ha_3}
\end{figure}

\emph{J02490228-1029220: } J024902 is a M1.5+M3.5+M3.5 resolved triple system for which we present three epochs of RV measurements. \citet{Bergfors2016} present an earlier FEROS spectrum of J024902, recording a RV of $17.1 \pm 1.1\ $ kms$^{-1}$ at MJD = 55901.121 along with a $v$ sin $i$ of $11\pm 3 \ $ kms$^{-1}$. \citet{Bergfors2016} also find J024902 to be a strong candidate member of the $\beta$ Pic moving group, based on UVW galactic velocities and spectroscopic age indicators such as strong Li absorption. However, a parallax measurement is required to confirm group membership. \citet{Bergfors2016} additionally derive M4.0 $\pm$ 1.0 spectral types for both BC components from resolved SINFONI \citep{Eisenhauer2003, Bonnet2004} spectra, consistent with the photometrically derived spectral types taken from \citet{Janson2012}.
\\
\\
\emph{J04244260-0647313: } J042442 is a previously known, three-component spectroscopic multiple, first reported in \citet{Shkolnik2010}. We recover a triple-peaked CCF and derive RV measurements for all three components in each epoch of observations. An example triple-peaked CCF is shown in Figure \ref{f:ccf_1}. Figure \ref{f:ha_3} shows the H$\alpha$ emission region of a single epoch, which is representative of the series of observations. J042442 displays strong H$\alpha$ emission at the predicted position given the measured RV of the primary.  \citet{Shkolnik2010} estimate M4.5, M5.5 and M5.7 $\pm\ 0.5$ component spectral types from HIRES \citep{Vogt1994} and ESPaDOnS \citep{Donati2006} spectra using the methods of \citet{Daemgen2007}. Using the spectral type - mass relations of \citet{Reid2005} they then estimate component masses of 0.17, 0.12 and 0.1 $M_{\sun}$ and place an upper limit of 1.9 days on the orbital periods of the BC pair and 70.3 days on the orbit of the BC pair around the primary. As with J053018 and  and J201633, we suspect that J042442 BC is a tidally synchronized pair as the 1.9 day orbital period is much shorter than the 10 day limit required for  synchronization to occur over $\sim$100~Myr timescales \citep[see e.g.,][]{Meibom2006}. However, unlike J053018 and J201633, J042442 displays strong H$\alpha$ emission at the predicted position given the measured RV of the primary. Therefore, J042442's X-ray emission is likely to be a genuine signature of youth, rather than due to tidal spin-up. Due to the tight nature of this system, $<0.25$ AU ($\lesssim7$ mas at 35 pc), it appears as a single star in our AstraLux images. Therefore, aside from possible interferometric applications, our derived RV measurements and planned future RV monitoring currently provide the only viable means to fully constrain orbital parameters and derive model independent component masses for this system.
\\
\\

\emph{J04373746-0229282 (GJ 3305): } GJ 3305 is a bona fide $\beta$ Pic moving group member \citep{Malo2013} and has a rich amount of astrometric data points, spanning $\sim15$ years of orbital motion. The binary is a wide \citep[$\sim2000$ AU, ][]{Feigelson2006} companion to the exoplanet host 51 Eri. It is a prime target for isochronal analysis that can be used to date the $\beta$ Pic group and test mass-luminosity evolutionary models. This has been accomplished by \citet{Montet2015} who exploit the wealth of astrometric information and a sample of complementary RV data from the literature to derive full orbital parameters and component masses to a good level of precision. However, the mass uncertainties are dominated by the uncertainty in RV semimajor amplitude, resulting from a lack of sufficient RV data points. 7 epochs of RV measurements for GJ 3305 are presented here, increasing the sample of available RV data points by a factor of $\sim2$. \citet{Montet2015} find BHAC15 evolutionary models \citep{Baraffe2015} to be consistent with individual component  masses to within 1.5$\sigma$ and derive a system age of 37 $\pm$ 9 Myr, consistent with the $\beta$ Pic age of 24 $\pm$ 3 Myr \citep[e.g.,][]{Bell2015, Mamajek2014}. This target is being densely monitored in both imaging and spectroscopy, and in future work, new orbital constraints will be deduced, with a considerable reduction in the error bars on the system age.
\\
\\

\emph{J07285137-3014490 (GJ 2060): } GJ 2060 is a bona fide AB Dor moving group member \citep{Malo2013} and has a large spread of astrometric data, sufficiently sampling and closing the binary orbit. In this study we record 10 new RV measurements for GJ 2060 which are plotted in Figure \ref{f:rv}. These measurements will significantly aid in orbital determinations and place tighter constraints on individual component masses. A detailed orbital analysis of this system and complementary spectral analysis of the components is underway (Rodet et al., submitted).
 \\
\\
\emph{J08475676-7854532 (EQ Cha): } EQ Cha is a member of the $\eta$ Cha association \citep{LopezM2013} and a suspected unresolved binary. Its binarity was first suspected due to its elevation in the Hertzsprung-Russell diagram \citep{Lawson2001, Luhman2004}. Surveys by \citet{Kohler2002} and \citet{Brandeker2006} then revealed partially resolved images of J084756, supporting the binary nature of this system. By fitting the elongated, partially resolved PSF profile \citet{Brandeker2006} measure a binary separation of 40 mas ($\sim4$ AU at 97 pc \citep{Bonavita2016}). This is within, or rapidly approaching, the diffraction limited resolution of the most advanced imaging instruments on 8m class telescopes. Therefore, again aside from possible interferometric applications, our derived RV measurements and future RV monitoring currently provide the only viable means to produce a firm detection of the binary and derive orbital parameters and component masses. 
 \\
\\
\emph{J12072738-3247002 (TWA 23): } TWA 23 is a bona fide member of the TW Hya association \citep{Malo2014} for which we report three epochs of RV measurements in Table \ref{tab:2}. We note that there exists two earlier epochs for this target listed in the FEROS archive which we additionally reduce, identifying two distinct peaks in the CCF. This suggests the system is a two-component spectroscopic binary and that the individual CCF peaks have merged in our later epochs due to the binary motion and spectrograph resolution. However, we note that the coordinates listed in the archive for the earlier epochs are offset by $\sim1.0\arcmin$ from the actual target coordinates. Systems within a $2\arcmin$ radius of TWA 23 are $>6$ magnitudes fainter, making it unlikely the target was misidentified within this region. However, \citet{Bailey2012} report RV measurements to a good level of precision ($\sim$60 ms$^{-1}$) for this system, identifying it as a single-lined spectroscopic binary from well sampled observations over the same timescale as the estimated orbital period, $\sim 4$ years. This suggests the early archival FEROS epochs, displaying a double peaked CCF, are not reliable. As we can not confidently conclude whether or not TWA 23 was indeed observed during these epochs, we exclude the RVs measured for the components from Table \ref{tab:2}, however we list them here for completeness; MJD = 54170.258, RV = -5.59 $\pm$ 0.22 and 19.10 $\pm$ 0.41\ kms$^{-1}$; MJD = 54228.241, RV = -3.51 $\pm$ 0.25 and 17.66 $\pm$ 0.41 kms$^{-1}$.
\\
\\
\emph{J15573430-2321123: } J155734 is a high probability member of the Upper Scorpius subgroup of the Sco-Cen association \citep{Rizzuto2015}. \citet{Kraus2008} resolved this system using aperture masking interferometry, deriving a binary separation of $\sim50$ mas and estimating a mass ratio of 0.59 $\pm\ 0.06$ from the measured contrast following the methods of \citet{Kraus2007}. \citet{Laf2014} partially resolve the system and use a template to fit the elongated PSF profile to measure binary separation and contrast, which they use to estimate a mass ratio of 0.39 with uncertainties on the order of 10 - 30\%. This system will 
notably benefit from our measurements and future RV analysis over a longer baseline, allowing the mass ratio discrepancy reported in the literature to be resolved and full orbital parameters to be derived for this tight binary.
\\
\\
\emph{J20531465-0221218: } J205314 is a potential member of the Argus moving group \citep{Malo2013} for which we report two epochs of RV measurements in Table \ref{tab:2}. As with TWA 23, we note there exists an additional archival epoch, which we reduce to recover a double peaked CCF,  suggesting the system is a two-component spectroscopic where the individual CCF components have merged during our later epochs. Again however, the coordinates listed are offset by $\sim1.0\arcmin$ from the actual target coordinates. As we can not confidently conclude whether or not J205314 was observed in this archival epoch, we treat this system in a similar fashion to TWA 23 and exclude the data from Table \ref{tab:2}, but list our measurements here for completeness; MJD = 56809.274, RV =  -45.32 $\pm$ 0.24 and -32.50 $\pm$ 0.30 kms$^{-1}$.
 \\
\\
\emph{J23205766-0147373: } J232057 is a high probability member of the Argus moving group \citep{Malo2014} and was resolved by \citet{Daemgen2007} with a separation of $\sim0.1\arcsec$. However, the system has since appeared as a single star in our AstraLux images, indicating that the binary companion has moved inward. Our RV measurements are of significant importance for this system as they provide the most viable means of sampling the orbital motion over this timeframe, until the two components become visible again.

\end{appendix}

\end{document}